# Microsphere-assisted generation of localized optical emitters in 2D hexagonal boron nitride

*Xiliang Yang,[†] Dong Hoon Shin,[±] Kenji Watanabe,[§] Takashi Taniguchi,[§] Peter G. Steeneken,[†] Sabina Caneva,[†]\**

[†] Department of Precision and Microsystems Engineering, Delft University of Technology, Mekelweg 2, 2628 CD, Delft, The Netherlands.

[±] Department of Electronics and Information Engineering, Korea University, Sejong 30019, Republic of Korea.

[§] National Institute for Materials Science, 1-1 Namiki, Tsukuba, Ibaraki 305-0044 Japan

\* Corresponding author. Email: s.caneva@tudelft.nl

ABSTRACT

Crystal defects in hexagonal boron nitride (hBN), are emerging as versatile nanoscale optical probes with a wide application profile, spanning the fields of nanophotonics, biosensing, bioimaging and quantum information processing. However, generating these crystal defects as reliable optical emitters remains challenging due to the need for deterministic defect placement and precise control of the emission area. Here, we demonstrate an approach that integrates microspheres (MS) with hBN optical probes to enhance both defect generation and optical signal readout. This technique harnesses MS to amplify light-matter interactions at the nanoscale through two mechanisms: focused femtosecond (fs) laser irradiation into a photonic nanojet for highly localized defect generation, and enhanced light collection via the whispering gallery mode effect. Our MS-assisted defect generation method reduces the emission area by a factor of 5 and increases the fluorescence collection efficiency by approximately 10 times compared to MS-free samples. These advancements in defect generation precision and signal collection efficiency open new possibilities for optical emitter manipulation in hBN, with potential applications in quantum technologies and nanoscale sensing.

**Introduction**

Nanoscale optical emitters are a cornerstone of nanophotonics, bioimaging and biosensing and an invaluable tool for investigating dynamics, mechanics, and interactions of biological components in aqueous environments.[1] While traditionally relying on fluorescent dyes for biophysics research, novel probes are emerging that offer better combinations of chemical and mechanical stability, brightness, and lifetime.[2] Among next-generation probes, 2D material optical emitters (OE) based on hexagonal boron nitride (hBN) crystal defects are particularly attractive.[3] These nanoscale probes exhibit exceptional brightness (4000 kcts/s),[4] long lifetimes (~3 ns),[5] high quantum efficiency (87%),[6] emission in the visible,[7] room-temperature operation,[8] stability in liquid environments,[9] and biocompatibility,[10] making them ideal for a wide range of applications. Unlike other 2D material-based emitters, such as transition metal dichalcogenides,[11] hBN defects are fully operational at room temperature, in liquids and in physiological conditions[12], significantly broadening their potential in both fundamental research and applications. However, the use of hBN emitters also face several challenges: 1) Fabrication can be complex, costly or low throughput requiring advanced techniques like ion/electron beam irradiation,[13, 14] femtosecond (fs) laser writing with high numerical aperture objective lenses or indentation with sharp atomic force microscopy (AFM) tips.[15] 2) Controlling the nature and location of defects within the crystal structure is difficult without modifying the substrate through the addition of e.g. nanopillars, wrinkles or microsphere arrays, impacting reproducibility and performance[16, 17] Addressing these challenges is crucial to integrate hBN emitters into high-resolution imaging and nanophotonics applications, necessitating further advances in fabrication and emission collection methods.

To generate OE in hBN, various fabrication methods have been developed, and are broadly categorized into two groups: methods with random spatial distributions and site-specific techniques.[17, 18] The former methods include thermal annealing,[19, 20] chemical[21] and plasma etching,[22] substrate strain-induced fabrication,[23] and bottom-up growth.[24] These techniques typically create defects over larger areas with little or no spatial control or require non-planar surfaces features.

In contrast, spatially controlled techniques, such as AFM indentation ion/electron beam irradiation, and direct laser writing, offer more precise control over emitter placement.[25, 26] However, they face restriction in terms of scalability and unwanted damage caused by the direct contact on the AFM tip,[25, 27, 28] or introducing foreign atomic species/implantation of ions.[29-31] Direct laser writing, especially using femto- to pico-second lasers, achieves precision and scalability,[32, 33] yet high laser powers can lead to considerable surface roughening, caused by thermal damage over the irradiated area.[34] Additionally, the spot size cannot be reduced below the diffraction limit determined by the numerical aperture (NA) of the objective lens, which restricts the spatial precision of the fabrication process.

Here, we report an approach utilizing a microsphere (MS) membrane to reduce the size of the irradiated area and concurrently control the position of optically generated hBN emitters. Leveraging the photonic jet focusing capability of MS, previously demonstrated for $SiO_2$/Si and gold surfaces,[35] this method achieves a fivefold reduction in the size of the area in which emitters are generated compared to fs-laser writing without MS, resulting in more localized hBN emitters. This improvement arises from the effective NA enhancement by the MS, which results in a focal spot size of less than a micron. Additionally, due to the principle of reversible optical paths, MS-assisted fabrication significantly boosts fluorescence collection efficiency by approximately 10 times. The enhancement of the optical absorption by the defects and reduced background noise[36, 37] is further accompanied by the enhanced the capture of refracted light through whispering gallery modes (WGM). In this process, light traveling around the MS's surface is confined by total internal reflection, creating resonant modes that amplify and direct the light. This effect optimizes the extraction of photons into the far field, improving the efficiency of light collection and emission from the optically-active hBN defect regions.[38] Practically, the MS traps the emitted light and directs it efficiently into a preferred out-coupling direction, further amplifying the fluorescence signal strength.

Therefore, MS integration not only enhances the spatial accuracy of emitter fabrication but also maximizes fluorescence signal collection. This platform can be readily integrated into single-molecule fluorescence microscopy systems, where hBN surfaces are finding increasing use as biocompatible substrates for wide-field imaging of biomolecules and their dynamics.[39] Additionally, by leveraging existing MS array technology,[40] and the simple integration in microfluidic devices, this MS implementation has the potential to achieve highly parallel signal collection in optofluidic sensing for health and environmental monitoring.

**Results and discussion**

**Microsphere-assisted laser fabrication.**

Figure 1a shows the laser incident on the MS, which forms a photonic nanojet (PNJ) upon emerging at the bottom of the MS. The femtosecond laser beam (515 nm, pulse duration 290 fs and linearly polarization) was focused through a Theta lens. To precisely control the MS position, we use high-refractive index (n=1.9) barium titanate ($BaTiO_3$) microspheres (diameter = 50 μm) embedded in PDMS membrane, which simplifies handling compared to controlling the position MSs without support. The MS-PDMS membrane (MPM) fabrication process is detailed in section S1 of the Supporting Information (SI). We note that while we focus here on single MS-based hBN emitter fabrication, self-assembly of MS during membrane embedding can generate tightly packed arrays that can increase the fabrication throughput. The MPM was placed on a $SiO_2$/Si substrate on which mechanically exfoliated, high-quality hBN flakes were transferred. To ensure the focus point could reach the hBN surface, we inserted a spacer consisting of double-sided tape between the MPM and the substrate. To determine the changes in optical field distribution caused by the MS, finite-difference time-domain (FDTD) simulation designs were conducted. The results show that when the MPM directly touched the substrate, it produced ring-shaped patterns due to focusing inside the MS, as shown in Figures 1b and 1c. Increasing the distance between the hBN and the MS improved PNJ focusing, as shown in Figure 1d and 1e, with a full-width at half-maximum (FWHM) of 480 nm at a distance of ~6 µm from the MS, which is smaller than the focal spot size of most high-NA objective lenses. Additionally, the PNJ length was approximately 5 µm, providing a large working distance tolerance for fabrication.

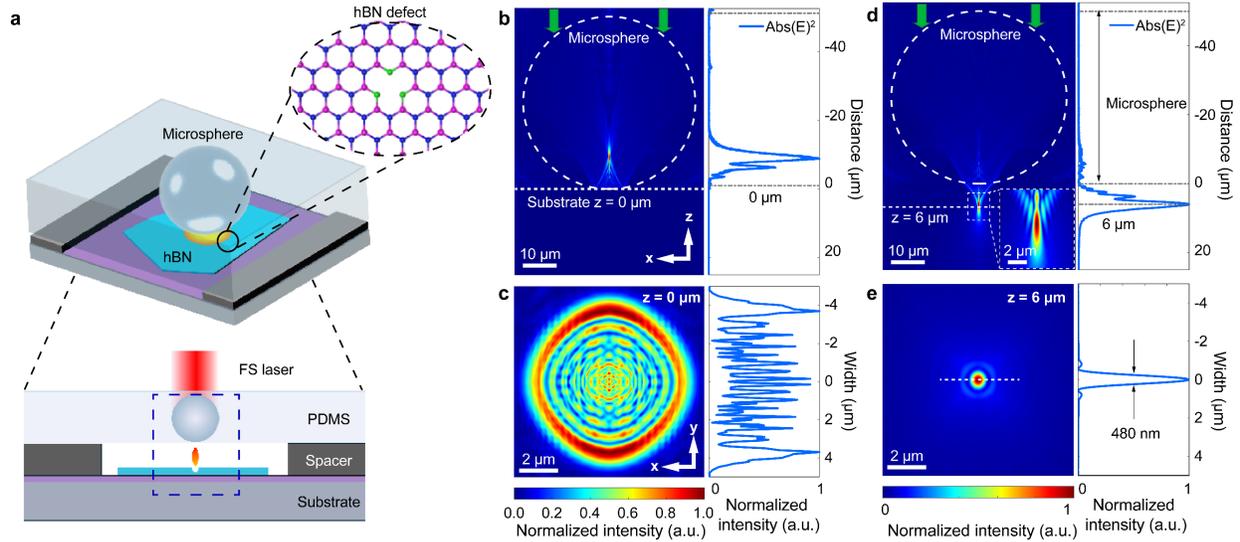

**Figure 1.** Schematic diagram of MS-assisted fs-laser fabrication of hBN emitters and simulations. **(a)** Experimental design of the Microsphere-PDMS Membrane (MPM) with spacer over the hBN flake surface on the SiO$_2$/Si substrate. Bottom: Side view of MS enhancement of the fs-laser focus. **(b)-(e)** FDTD simulation of the light field distribution of the MPM focus: **(b)** Light distribution in the xz-plane and intensity distribution along the z-axis with direct contact between the MPM and the substrate. **(c)** Light distribution in the xy-plane and intensity distribution along the y-axis at the maximum intensity position with direct contact. **(d)** Light distribution in the xz-plane and intensity distribution along the z-axis with a 6 µm distance between hBN and MS. **(e)** Light distribution in the xy-plane and intensity distribution along the y-axis at the maximum intensity position with a 6 µm distance.

Based on the stack design and simulations in Figure 1, fs-laser fabrication was used to generate defects in hBN using a 6-µm spacer, the profile of which is depicted in Figure S2. The irradiation spots on the xy-plane were spaced approximately 30 µm apart to avoid crosstalk. Figure 2a compares the results of femtosecond laser processing without and with microspheres. Area I shows that during direct fs-laser writing, the hBN layer (yellow) is fractured, exposing the $SiO_2$/Si substrate (blue) underneath it. When the laser pulse is applied directly to the $SiO_2$/Si, holes are formed in the surface (black). Area II shows MPM-assisted laser writing, which results in the formation of an hBN bubble. Without the MPM layer, the fs-laser directly irradiates the hBN surface, leading to extensive breaking of the lattice, as shown in area I in Figure 2a, producing an irregular opening of ~15 µm. Strikingly, on the same flake, laser writing with the MPM layer reduces the affected area (area II) to a diameter of approximately 3.6 µm. This reduction in diameter corresponds to a decrease in the affected area by a factor of ~5. The bubble exhibits a circular shape, with a radius of 1.8 µm and a height of 45 nm (h/R = 0.025), as shown by the AFM analysis (Figure 2b,c). This observation suggests that a tightly focused PNJ causes deformation of the top layers of the hBN flake ($t_{hbn}$ ~50 nm), caused by localized heating from high-intensity laser pulses, which leads to thermal expansion and blistering.[35] Bubble formation in 2D materials has been previously attributed to mechanical stress from rapid heating and cooling, which can trap air as well as ambient, or surface absorbants.[41, 42] The bubble's ratio h/R=0.025 here is lower than that of monolayer hBN bubble formation (h/R=0.11),[43] but is in good agreement with the multilayer hBN bubble formation seen in both experimental and theoretical results.[42]

Figure 2e shows the photoluminescence (PL) of the hBN bubble, with most emitters located at the edges of bubble, where the strain is most pronounced. This suggests that the emission is largely dependent on lattice strain, as evidenced by a Raman hBN peak shift from 1365 to 1363 cm$^{-1}$ along the bubble, indicating tension in the hBN flake.[42] Similar peak shifts due to maximized strain at the bubble's center have previously been reported.[42] Figure 3f displays the PL spectra from these bright spots (circles 1 and 2 in Figure 2e), showing the typical zero-phonon lines (ZPL) of hBN emitters.

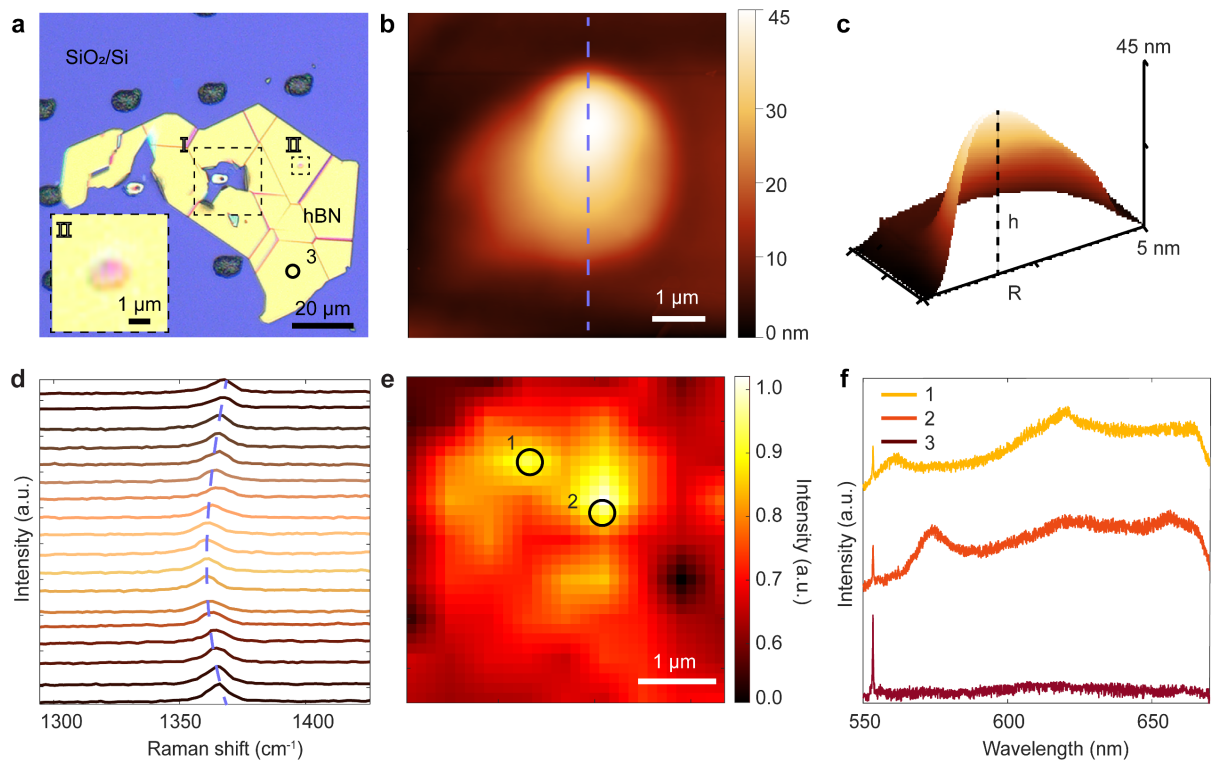

**Figure 2**. Characterization of fs-laser induced hBN defects. **(a)** Optical image of the hBN surface after fs-laser irradiation with (area II) and without (area I) MPM (inset: zoomed view of the MPM irradiated area, scale bar: 1 µm). **(b)** 2D AFM image of the hBN bubble. **(c)** 3D AFM image of the bubble cross section. **(d)** Raman spectra along the diameter of the bubble (highlighted in (b) by a purple dashed line). **(e)** PL mapping of the hBN bubble. **(f)** Emission spectra of two representative emitters and background, labeled as circles 1 and 2 in (e), and 3 in (a), respectively.

Figure 3 demonstrates different sizes of defects produced with MPM by varying the laser irradiation power. When the power was below the breakdown threshold (~0.07 W) in our setup, slight color changes are visible in the hBN, but no clear defect areas could be observed. However, at higher power levels, clear visible structural changes appeared in the material. Details of the laser-induced defects are shown in Figure 3a and Figure S3.

When the laser power exceeded the breakdown threshold but remained below 0.3 W, bubble patterns were observed. The diameter of these bubble structures ranged from 3-4 µm, which was significantly smaller than the focal spot size of the f-theta lens (9 µm at $1/e^2$ intensity), as shown in Figure 2a , 3a and S3. This size reduction can be attributed to the PNJ generated by the MS, which effectively reduces the focal spot size. Pulses with power above 0.3 W caused ablation at the center of the PNJ, resulting in the outward redeposition of hBN material around the hole. With 0.3 W and 0.45 W laser power, the defect patterns generated by MPM enlarged to 4.5 µm and 5.6 µm, respectively, breaking up into holes. In comparison, direct laser fabrication patterns were about 5 times larger than those produced by the MPM method as shown in Figure S3. The defect patterns exhibited torn edges, likely formed by the high pressures and shock waves generated by the high-energy laser pulses, accompanied by a breakdown of the surrounding material, similar to laser-induced micro explosions confined within the bulk of sapphire.[44] Additionally, the high concentration of heat could induce oxidation of the silicon surface, as previously described.[35] The average area of the defect patterns, based on over 15 samples, increased linearly with the pulse power with a scaling factor of 13 µm/W, as shown in Figure S3e.

We note that the observed variation in the size of the defect patterns, even with the same pulse energies, can be attributed to factors such as the variation in hBN flake thickness ($t_{hBN}$: 40 nm - 70 nm), the concentration of impurities or defects in the bulk material of hBN, wrinkles formed during exfoliation, and misalignment of the laser writing setup and the microsphere setup.[15, 45] These factors also prevent the formation of smaller defects as predicted by simulations.

To activate and stabilize the photoemission from the hBN defects and remove the contamination induced during the fabrication process, the hBN samples were annealed in 1000 °C at $10^{-7}$ bar for 2 hours after laser writing with the MPM. As previously reported, annealing plays an important role in restructuring and forming new optically active colour centres in hBN.[33] Before annealing, few PL signal was observed from the laser-processed sample. After annealing, sharp, bright PL peaks appeared at laser-irradiated sites. Unlike emitters produced by conventional direct thermal annealing methods, which are located on randomly generated wrinkles due to the differing thermal expansion coefficients of hBN and the substrate.[20] PL maps were subsequently obtained on these processed areas using a 0.85 NA objective with a 514 nm wavelength continuous-wave laser for excitation, (Figure 3b). The maps indicate that multiple emission centers are generated and distributed along the edge of the affected areas. Moreover, the number and type of emitters changed with the defect pattern size. Single and double emission spots can be seen around the defect pattern in the bubble structure, yet due to the diffraction limit of the optical system, we cannot identify and determine the number of individual nanoscale emitters. When the pattern size exceeded ~3 μm, the emission spots started to merge into clusters, as shown in the middle panel of Figure 3b. When the fabrication laser power was increased to 0.45 W, the hBN surface underwent significant damage, with folded edges, and thus with emitters likely located at different heights from the surface. The strongest emission peaks, centered at 560, 600, 620, and 626 nm, match the ZPL of previously reported hBN emitters (Figure 3c).[39] Additionally, sharp Raman peaks of hBN (located at 553 nm) also appear in the spectra. Figure 3d shows histograms of the ZPL wavelengths for all stable emitters across the three defect pattern groups. Based on their ZPL color distribution, the emitters are categorized into four groups: green (555 ± 15 nm), yellow (580

± 10 nm), orange (605 ± 15 nm), and red (650 ± 30 nm). For emitters characterized by the representative bubble structure, a total of 12 samples were analyzed, with the zero-phonon line (ZPL) peak distribution predominantly falling within the orange and red emitter ranges. As the defect size increases (i.e. increase in laser power), the bubble structures transition to holes, where the strongest peak distribution shifts from longer wavelengths (600-640 nm) to shorter wavelengths (550-590 nm), and very few emitters exhibit ZPLs longer than 600 nm, as shown by the percentage distribution in Figure S4. This is in agreement with previous observations in which fs-fabrication of hBN at laser energies insufficient to cause ablation, resulted in bubble-like hBN defects instead of holes.[12] Furthermore, the ZPL full width at half-maximum (FWHM) also decreases by 1.1 nm as the laser pulse energy increases from 0.15 W to 0.45 W. For defect patterns obtained with 0.45 W irradiation, the ZPLs are narrow and sharp, comparable to other emitter fabrication methods (e.g. FIB, AFM tip).[14, 25]

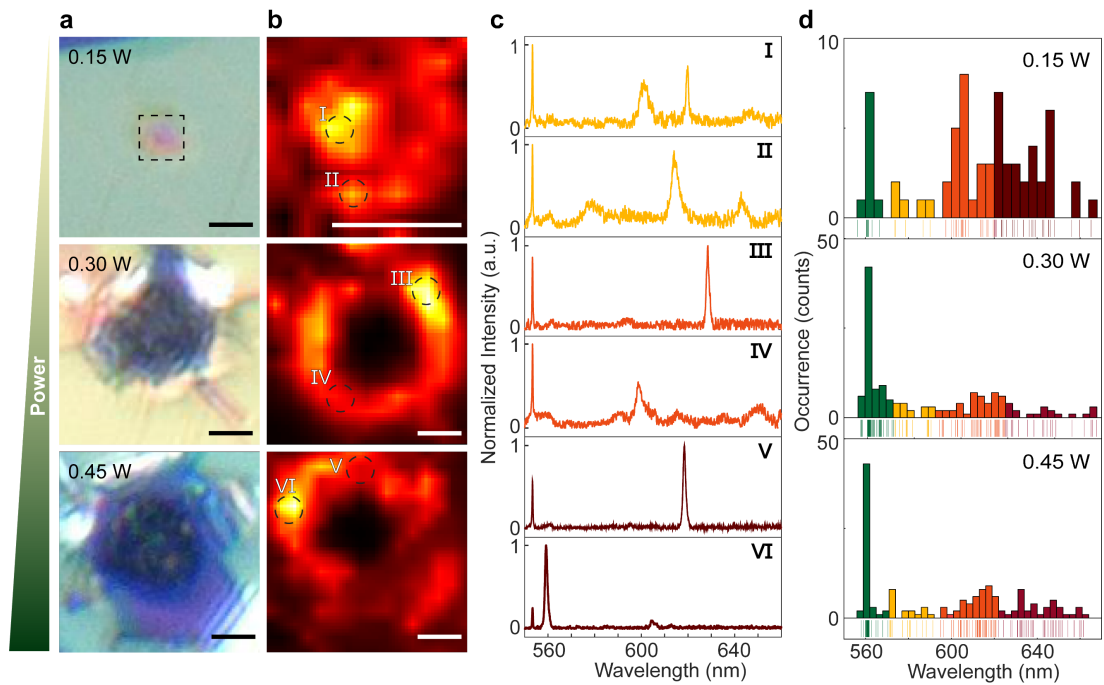

**Figure 3 (a)** Optical images of the defects of different sizes generated by MPM fs-laser with powers of 0.15, 0.3, 0.45 W. Scale bar: 3 μm. **(b)** The corresponding PL maps of the irradiated areas in (a). Scale bar: 3 μm. **(c)** Emission spectra of six representative emitters, as labelled in (b).

The 550 nm peak in all panels arises from the Raman contribution. **(d)** Histograms of the emitters ZPLs, categorized by different colour ranges. Bin size: 3 nm.

**Microsphere-enhanced optical signal collection.**

Beyond using microspheres to deterministically generate hBN optical emitters with a tightly focused PNJ and reduced irradiated area, we further demonstrate that the same MPM layer can be employed as a versatile component for enhancing optical signals. As shown in Figures 4, the MPM significantly boosts the intensity of both excitation and emission signals. This is not solely due to the focusing effect, but results from two key factors: the efficient coupling of both near-field and far-field energy from the hBN optical emitters, and the additional enhancement provided by whispering gallery modes (WGM), which together improve the transmission to the subsequent optical components. This enhancement allows optical emitters to be imaged with low NA objectives (NA=0.6, 50x) without requiring an oil-immersion high NA lens (NA>1), enabling high-resolution imaging with a simpler, lower-cost optical setup.

The incorporation of the MPM modifies the transmission and distribution of light emitted by the optical emitters (Figure 4a,b), resulting in notable enhancements in signal intensity and imaging resolution. These improvements are driven by two key mechanisms:

1) Spatial localization of the excitation area: The MPM confines the laser focus to the submicron scale (Figure 1g), concentrating energy on a smaller, well-defined three-dimensional region. This enhanced focus increases absorption efficiency at defect sites while minimizing excitation in non-target areas, thereby reducing background noise and preventing unwanted excitation of non-relevant emitters. Consequently, the overall excitation efficiency of the desired emitters is significantly improved.[46, 47]

2) Enhancement of the signal extraction efficiency: When emitters are excited outside the plane of the substrate and directly coupled into the MPM, the light is refracted and enhanced by whispering gallery modes (Figure 4c), significantly boosting the extraction rate of generated photons into the far-field. COMSOL simulations indicate this enhancement exceeds 4000 times at the top of the microsphere, as shown in Table S6. This mechanism optimizes the propagation path of photons, improving their output efficiency and enhancing the fluorescence signal's intensity and stability.[23]

Through the synergistic effect of these mechanisms, the MPM can be used to significantly enhance the optical detection of hBN emitters. This system design allows us to achieve efficient imaging of hBN emitters without using high NA objectives, further advancing the development of low-cost, high-performance optical emitter imaging applications.

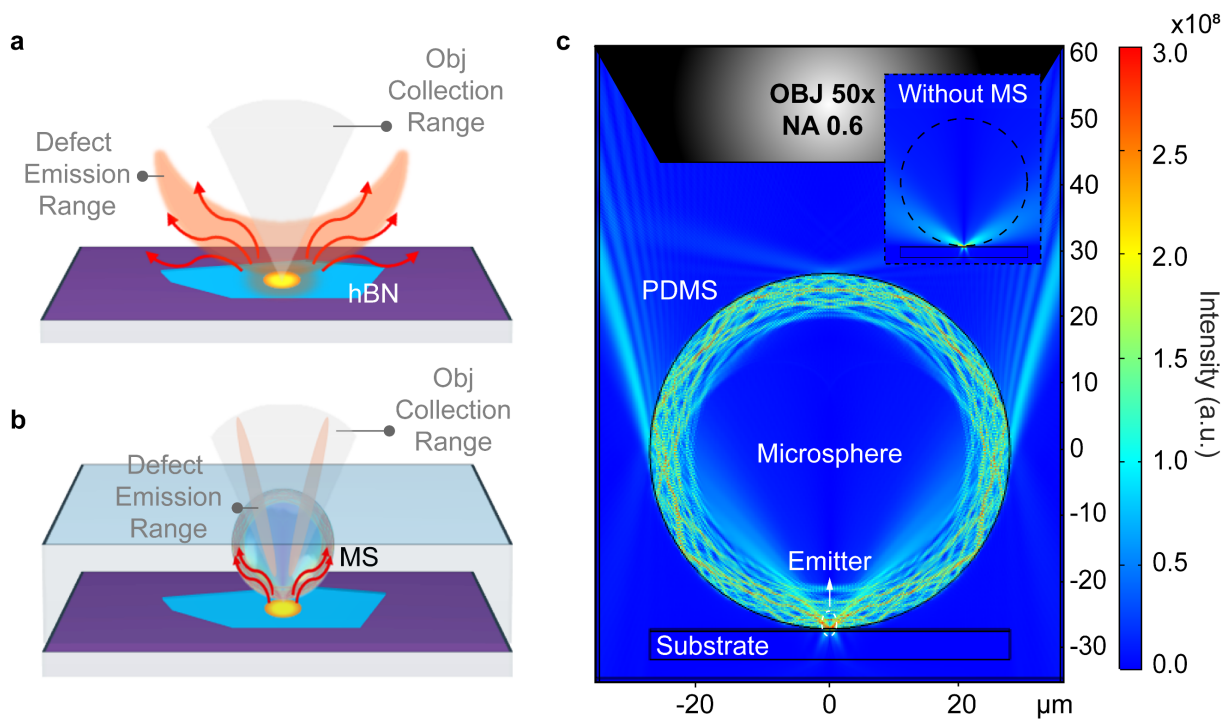

**Figure 4.** Schematic of MS-enhanced fluorescence collection. **(a)** Illustration of the spatial emission distribution of hBN emitters without MS. The orange area shows the defect emission range, and the red arrows indicate the extreme angle of defect emission (we note that it will be in 3D). **(b)** Illustration of the spatial emission distribution of hBN emitters with a MS, where emitted photons are coupled to the microsphere via whispering gallery modes and re-radiated at an engineered angle that falls within the collection range of the objective lens (light grey area). **(c)** Simulation of optical whispering-gallery modes in the MPM system, enhancing the fluorescence signal of the emitters. The inset shows the spatial emission distribution of an emitter without an MS.

The MPM with a single MS was first employed to quantitatively estimate the enhancement ratio by enhancing the hBN Raman peak. This measurement requires less localization since the Raman peak can be detected anywhere on the hBN flake surface. A laser beam was focused using a 50x objective with an NA of 0.6 onto the center of the microsphere to achieve maximum enhancement. Figure 5a shows the Raman spectra obtained using MPM with different gaps: 1 µm spacer with MS (1µm-SP-MS), 0.5 µm spacer with MS (0.5 µm-SP-MS), and 0 µm spacer with MS (0µm-SP-MS). Two control groups were also included: one with only PDMS and another uncovered hBN. With two control groups, we can determine the original intensity of the Raman peak on the flat hBN surface, and the peak contributed by PDMS.

For MPM without a gap, the hBN Raman peak showed approximately a factor 20 enhancement compared to the situation without MPM, but the enhancement factor decreased to ~4 and ~2 for 0.5 µm and 1 µm gaps, respectively. This demonstrates that most of the photon energy is lost in free space without MS-enhanced collection within these short distances. This also clarifies why, despite simulations predicting a 4000-fold enhancement, real experiments achieve lower values, largely due to the challenge of perfectly aligning the emitter at the exact center of the microsphere's focal region (Figure S5b, points A and B). As a result, some enhancement is lost. Thus, achieving close attachment of the MPM to the surface is crucial for maximizing emission collection efficiency.

We note that while the same MPM chips can be used for both fabrication and collection at an intermediate distance of 0 and 6 μm between the microsphere and the substrate, the performance will be slightly compromised due to this gap size. As shown in this study, varying this gap can optimize performance depending on the desired application, providing flexibility in choosing the most suitable gap size for either extraction or collection.

Next, the MPM was attached directly to an emitter at the hBN flake edge, created via MS fs-laser writing. The MS serves as a small lens, forming a virtual image of the sample surface,[48] allowing us to precisely localize the emitters at the edge (inset, Figure 5b; optical image in Figure S6). This ensures the MPM overlaps with and enhances emission from the region of interest.

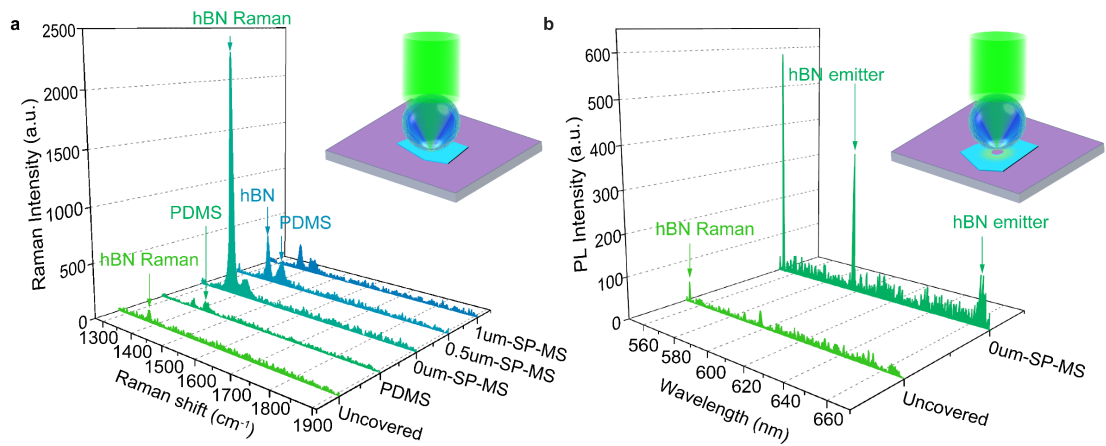

**Figure 5.** Signal enhancement from hBN flakes covered with MPM. **(a)** Raman spectra of an hBN flake enhanced by MPM with varying spacer distances. Inset: Schematic of the Raman measurement setup for the hBN flake covered by MPM. **(b)** PL spectra of hBN flake emitters fabricated by MPM fs-laser, comparing MPM-covered versus uncovered conditions. Inset: Schematic of the PL measurement setup for emitters at the edge of the hole with the MPM in place.

The MPM is mainly composed of $BaTiO_3$, which has a wide bandgap energy (~3.2 eV) and does not support low-energy excitons, resulting in negligible optical loss in the visible wavelength band.[49] This property makes $BaTiO_3$ an excellent optical dielectric, enabling efficient light transmission and manipulation. In Figure 5b, the PL spectra of the hBN emitter collected with (dark green) and without (light green) the MPM show a significant enhancement at 598 nm, with the MPM amplifying the PL signal by approximately a factor of 10.

However, the PL peak enhancement ratio is lower compared to the Raman peak. This can likely be attributed to BaTiO$_3$'s limited interaction with low-energy excitons, which reduces its effect on PL signals at lower energies despite minimizing optical losses. Additionally, slight spatial misalignment between the emitter and the BaTiO$_3$ microsphere could create a gap as shown in Figure S5b, allowing some surface waves, such as evanescent waves, to be lost into space rather than being collected by the microsphere and subsequent optical systems. These factors may contribute to the reduced enhancement. Despite these factors, the observed PL enhancement demonstrates the effectiveness of the MPM in improving signal strength, making full use of BaTiO$_3$'s favourable optical properties.

**Conclusion**

In summary, our study demonstrates a new route enabling the combined fabrication of hBN emitters and fluorescence emission enhancement by using a MS chip as an effective and low-cost focusing lens. By combining a MS embedded in a PDMS membrane (MPM) with a fs-laser writer setup, we achieve a fivefold reduction in the irradiated area, leading to better localization and higher quality emitters (i.e. smaller FWHM of the ZPLs) in hBN, and suppressing the extensive damage found in MS-free fabrication. This approach not only enables better control over the hBN defect generation process but also substantially improves optical signal collection efficiency by approximately 10 times compared to MS-free measurements methods. The enhancement in defect absorption, combined with optimized photon extraction and efficient light directionality imposed by the MS geometry, results in significantly stronger optical signal detection. We emphasize that this proof-of-concept emitter fabrication and emission collection with MS can be highly parallelized through the use of self-assembled MS arrays over large-area hBN surfaces. The MPM can also be integrated in microfluidics, enhancing the detection of fluorescent biomolecules in physiological conditions. Furthermore, our findings can be readily applied to other 2D materials exhibiting optically-active defects, setting the stage for further developments in nanophotonics and fluorescence imaging at 2D material surfaces.

**Methods**

**Sample Preparation and Laser Irradiation**

Multilayer hBN flakes were first mechanically exfoliated from bulk hBN crystals produced by high temperature and high-pressure synthesis (NIMS Japan) and then transferred onto $SiO_2$/Si surface cleaned by ultrasonic sonication in acetone and IPA for 3 mins and oxygen plasma cleaning for 5 mins. Using atomic force microscopy (AFM, Cypher), large area hBN thin flakes with the thickness of ~50 nm were selected. We use high-refractive index (n=1.9) barium titanate microspheres (diameter = 50 μm, Cospheric, USA) embedded in PDMS (SYLGARD 184 Silicone Elastomer Kit, Germany) to generate a microsphere-PDMS membrane (MPM). we inserted a ~6 μm spacer consisting of double-sided tape (Zhuanyi Electronic Elechnology Co., SuZhou, China) between the MPM and the substrate.

   Subsequently, a single fs-laser pulse (Pharos PH1-15, Light Conversion) of wavelength 515 wavelength, pulse duration 290 fs and linearly polarization was focused through a f-theta lens (F=100 mm, 510-550 nm Fused Silica Telecentric, VONJAN). The objective lens was mounted on a piezo nano positioning stage (PI E665) with nanometer resolution to precisely control the focal position. After laser irradiation, the hBN samples were annealed in a tube furnace at 1000 °C in air for 1h.

**Optical measurements**

Raman and PL spectra were acquired in Renishaw Raman setup. The hBN samples were excited by a 514 nm argon laser (MODU-LASER) with 0.15 mW. The spectra were collected by microscope objective lenses (Olympus 50×) with numerical aperture (NA) of 0.6 and a 1800 l/mm grating. The PL mapping was performed with the same setup and excitation laser, but with higher resolution with objective (Olympus 100x) with numerical NA 0.9, except for the laser scanning step of 0.5 μm and 1 μm.

**Photonics Simulation**

Numerical simulations of the electromagnetic field were conducted using both and a finite-difference time-domain (FDTD) method in Ansys Lumerical FDTD for focusing analysis and a finite element method (FEM) in COMSOL Multiphysics for optical WGMs. The simulations were performed over an area of 60 μm × 30 μm with a mesh size of $\lambda/100$. Perfectly matched layers and periodic boundary conditions were applied to ensure accuracy, and the boundary matching layers were set as perfect absorption layers. The refractive indices of the PDMS film and microspheres (diameter = 50 μm) were set to 1.4 and 1.9, respectively, with the incident light as a plane wave at 514 nm wavelength. The simulation conditions were consistent with those of the experiment.

ASSOCIATED CONTENT

**Supporting Information**

The following files are available free of charge.

S1. Fabrication of the microsphere-PDMS membrane film.

S2. Height profile along the edge of the tape spacer.

S3. Optical images of defects created by direct fs laser fabrication.

S4. Percentage distribution of ZPL wavelengths with increasing fabrication power.

S5. Simulated electric field distribution for the emitter collection.

S6. Simulated electric field distribution on the microsphere.

S7. Optical images at different image planes of the microsphere.

## AUTHOR INFORMATION

**Corresponding Author**

* Email: s.caneva@tudelft.nl

**Author Contributions**

The manuscript was written through contributions of all authors. All authors have given approval to the final version of the manuscript.

ACKNOWLEDGMENT

X.Y. acknowledges funding from the Chinese Scholarship Council (Scholarship No. 202108270002). S.C. acknowledges funding from the European Union's Horizon 2020 research and innovation program (ERC StG, SIMPHONICS, Project No. 101041486). All authors acknowledge K. Watanabe and T. Taniguchi from the National Institute of Materials Science (NIMS) for the bulk hBN crystals. We sincerely thank Binbin Zhang for his assistance with the whispering gallery mode simulations, and Zhenyuan Lin for his insights on fs-laser fabrication.

**Notes**

The authors declare no competing financial interests.

REFERENCES

(1) Fernandez-Suarez, M.; Ting, A. Y. Fluorescent probes for super-resolution imaging in living cells. *Nat Rev Mol Cell Biol* **2008**, *9* (12), 929-943. DOI: 10.1038/nrm2531

(2) Yang, X.; Zhang, Q.; Zhang, S.; Lai, M.; Ji, X.; Ye, Y.; Li, H.; Zhao, M. Molecule fluorescent probes for sensing and imaging analytes in plants: Developments and challenges. *Coordination Chemistry Reviews* **2023**, *487*. DOI: 10.1016/j.ccr.2023.215154

(3) Steinleitner, P.; Merkl, P.; Nagler, P.; Mornhinweg, J.; Schuller, C.; Korn, T.; Chernikov, A.; Huber, R. Direct Observation of Ultrafast Exciton Formation in a Monolayer of WSe(2). *Nano Lett* **2017**, *17* (3), 1455-1460. DOI: 10.1021/acs.nanolett.6b04422

(4) Martínez, L. J.; Pelini, T.; Waselowski, V.; Maze, J. R.; Gil, B.; Cassabois, G.; Jacques, V. Efficient single photon emission from a high-purity hexagonal boron nitride crystal. *Physical Review B* **2016**, *94* (12). DOI: 10.1103/PhysRevB.94.121405

(5) Tran, T. T.; Bray, K.; Ford, M. J.; Toth, M.; Aharonovich, I. Quantum emission from hexagonal boron nitride monolayers. *Nat Nanotechnol* **2016**, *11* (1), 37-41. DOI: 10.1038/nnano.2015.242

(6) Nikolay, N.; Mendelson, N.; Özelci, E.; Sontheimer, B.; Böhm, F.; Kewes, G.; Toth, M.; Aharonovich, I.; Benson, O. Direct measurement of quantum efficiency of single-photon emitters in hexagonal boron nitride. *Optica* **2019**, *6* (8). DOI: 10.1364/optica.6.001084

(7) Hayee, F.; Yu, L.; Zhang, J. L.; Ciccarino, C. J.; Nguyen, M.; Marshall, A. F.; Aharonovich, I.; Vuckovic, J.; Narang, P.; Heinz, T. F.; et al. Revealing multiple classes of stable quantum emitters in hexagonal boron nitride with correlated optical and electron microscopy. *Nat Mater* **2020**, *19* (5), 534-539. DOI: 10.1038/s41563-020-0616-9

(8) Grosso, G.; Moon, H.; Lienhard, B.; Ali, S.; Efetov, D. K.; Furchi, M. M.; Jarillo-Herrero, P.; Ford, M. J.; Aharonovich, I.; Englund, D. Tunable and high-purity room temperature single-photon emission from atomic defects in hexagonal boron nitride. *Nat Commun* **2017**, *8* (1), 705. DOI: 10.1038/s41467-017-00810-2

(9) Ronceray, N.; You, Y.; Glushkov, E.; Lihter, M.; Rehl, B.; Chen, T. H.; Nam, G. H.; Borza, F.; Watanabe, K.; Taniguchi, T.; et al. Liquid-activated quantum emission from pristine hexagonal boron nitride for nanofluidic sensing. *Nat Mater* **2023**, *22* (10), 1236-1242. DOI: 10.1038/s41563-023-01658-2

(10) Merlo, A.; Mokkapati, V.; Pandit, S.; Mijakovic, I. Boron nitride nanomaterials: biocompatibility and bio-applications. *Biomater Sci* **2018**, *6* (9), 2298-2311. DOI: 10.1039/c8bm00516h

(11) Zheng, W.; Jiang, Y.; Hu, X.; Li, H.; Zeng, Z.; Wang, X.; Pan, A. Light Emission Properties of 2D Transition Metal Dichalcogenides: Fundamentals and Applications. *Advanced Optical Materials* **2018**, *6* (21). DOI: 10.1002/adom.201800420

(12) Gan, L.; Zhang, D.; Zhang, R.; Zhang, Q.; Sun, H.; Li, Y.; Ning, C. Z. Large-Scale, High-Yield Laser Fabrication of Bright and Pure Single-Photon Emitters at Room Temperature in Hexagonal Boron Nitride. *ACS Nano* **2022**, *16* (9), 14254-14261. DOI: 10.1021/acsnano.2c04386

(13) Kianinia, M.; White, S.; Fröch, J. E.; Bradac, C.; Aharonovich, I. Generation of Spin Defects in Hexagonal Boron Nitride. *ACS Photonics* **2020**, *7* (8), 2147-2152. DOI: 10.1021/acsphotonics.0c00614

(14) Fournier, C.; Plaud, A.; Roux, S.; Pierret, A.; Rosticher, M.; Watanabe, K.; Taniguchi, T.; Buil, S.; Quelin, X.; Barjon, J.; et al. Position-controlled quantum emitters with reproducible emission wavelength in hexagonal boron nitride. *Nat Commun* **2021**, *12* (1), 3779. DOI: 10.1038/s41467-021-24019-6

(15) Gao, X.; Pandey, S.; Kianinia, M.; Ahn, J.; Ju, P.; Aharonovich, I.; Shivaram, N.; Li, T. Femtosecond Laser Writing of Spin Defects in Hexagonal Boron Nitride. *ACS Photonics* **2021**, *8* (4), 994-1000. DOI: 10.1021/acsphotonics.0c01847

(16) Chen, X.; Yue, X.; Zhang, L.; Xu, X.; Liu, F.; Feng, M.; Hu, Z.; Yan, Y.; Scheuer, J.; Fu, X. Activated Single Photon Emitters And Enhanced Deep‐Level Emissions in Hexagonal Boron Nitride Strain Crystal. *Advanced Functional Materials* **2023**, *34* (1). DOI: 10.1002/adfm.202306128

(17) Aharonovich, I.; Tetienne, J. P.; Toth, M. Quantum Emitters in Hexagonal Boron Nitride. *Nano Lett* **2022**, *22* (23), 9227-9235. DOI: 10.1021/acs.nanolett.2c03743

(18) Zhang, C.; Shi, Z.; Wu, T.; Xie, X. Microstructure Engineering of Hexagonal Boron Nitride for Single‐Photon Emitter Applications. *Advanced Optical Materials* **2022**, *10* (17). DOI: 10.1002/adom.202200207

(19) Li, C.; Xu, Z.-Q.; Mendelson, N.; Kianinia, M.; Toth, M.; Aharonovich, I. Purification of single-photon emission from hBN using post-processing treatments. *Nanophotonics* **2019**, *8* (11), 2049-2055. DOI: 10.1515/nanoph-2019-0099

(20) Yim, D.; Yu, M.; Noh, G.; Lee, J.; Seo, H. Polarization and Localization of Single-Photon Emitters in Hexagonal Boron Nitride Wrinkles. *ACS Applied Materials & Interfaces* **2020**, *12* (32), 36362-36369. DOI: 10.1021/acsami.0c09740

(21) Mendelson, N.; Morales‐Inostroza, L.; Li, C.; Ritika, R.; Nguyen, M. A. P.; Loyola‐Echeverria, J.; Kim, S.; Götzinger, S.; Toth, M.; Aharonovich, I. Grain Dependent Growth of Bright Quantum Emitters in Hexagonal Boron Nitride. *Advanced Optical Materials* **2020**, *9* (1). DOI: 10.1002/adom.202001271

(22) Xu, Z. Q.; Elbadawi, C.; Tran, T. T.; Kianinia, M.; Li, X.; Liu, D.; Hoffman, T. B.; Nguyen, M.; Kim, S.; Edgar, J. H.; et al. Single photon emission from plasma treated 2D hexagonal boron nitride. *Nanoscale* **2018**, *10* (17), 7957-7965. DOI: 10.1039/c7nr08222c

(23) Stewart, J. C.; Fan, Y.; Danial, J. S. H.; Goetz, A.; Prasad, A. S.; Burton, O. J.; Alexander-Webber, J. A.; Lee, S. F.; Skoff, S. M.; Babenko, V.; et al. Quantum Emitter Localization in Layer-Engineered Hexagonal Boron Nitride. *ACS Nano* **2021**, *15* (8), 13591-13603. DOI: 10.1021/acsnano.1c04467

(24) Ma, K. Y.; Kim, M.; Shin, H. S. Large-Area Hexagonal Boron Nitride Layers by Chemical Vapor Deposition: Growth and Applications for Substrates, Encapsulation, and Membranes. *Accounts of Materials Research* **2022**, *3* (7), 748-760. DOI: 10.1021/accountsmr.2c00061

(25) Xu, X.; Martin, Z. O.; Sychev, D.; Lagutchev, A. S.; Chen, Y. P.; Taniguchi, T.; Watanabe, K.; Shalaev, V. M.; Boltasseva, A. Creating Quantum Emitters in Hexagonal Boron Nitride Deterministically on Chip-Compatible Substrates. *Nano Lett* **2021**, *21* (19), 8182-8189. DOI: 10.1021/acs.nanolett.1c02640

(26) Hou, S.; Birowosuto, M. D.; Umar, S.; Anicet, M. A.; Tay, R. Y.; Coquet, P.; Tay, B. K.; Wang, H.; Teo, E. H. T. Localized emission from laser-irradiated defects in 2D hexagonal boron nitride. *2D Materials* **2017**, *5* (1). DOI: 10.1088/2053-1583/aa8e61

(27) Blundo, E.; Surrente, A.; Spirito, D.; Pettinari, G.; Yildirim, T.; Chavarin, C. A.; Baldassarre, L.; Felici, M.; Polimeni, A. Vibrational Properties in Highly Strained Hexagonal Boron Nitride Bubbles. *Nano Lett* **2022**, *22* (4), 1525-1533. DOI: 10.1021/acs.nanolett.1c04197

(28) Chen, L.; Elibol, K.; Cai, H.; Jiang, C.; Shi, W.; Chen, C.; Wang, H. S.; Wang, X.; Mu, X.; Li, C.; et al. Direct observation of layer-stacking and oriented wrinkles in multilayer hexagonal boron nitride. *2D Materials* **2021**, *8* (2). DOI: 10.1088/2053-1583/abd41e

(29) Mendelson, N.; Chugh, D.; Reimers, J. R.; Cheng, T. S.; Gottscholl, A.; Long, H.; Mellor, C. J.; Zettl, A.; Dyakonov, V.; Beton, P. H.; et al. Identifying carbon as the source of visible single-photon emission from hexagonal boron nitride. *Nat Mater* **2021**, *20* (3), 321-328. DOI: 10.1038/s41563-020-00850-y

(30) Glushkov, E.; Macha, M.; Rath, E.; Navikas, V.; Ronceray, N.; Cheon, C. Y.; Ahmed, A.; Avsar, A.; Watanabe, K.; Taniguchi, T.; et al. Engineering Optically Active Defects in Hexagonal Boron Nitride Using Focused Ion Beam and Water. *ACS Nano* **2022**, *16* (3), 3695-3703. DOI: 10.1021/acsnano.1c07086

(31) Ngoc My Duong, H.; Nguyen, M. A. P.; Kianinia, M.; Ohshima, T.; Abe, H.; Watanabe, K.; Taniguchi, T.; Edgar, J. H.; Aharonovich, I.; Toth, M. Effects of High-Energy Electron Irradiation on Quantum Emitters in Hexagonal Boron Nitride. *ACS Appl Mater Interfaces* **2018**, *10* (29), 24886-24891. DOI: 10.1021/acsami.8b07506

(32) Ahmad, S. I.; Dave, A.; Sarpong, E.; Yao, H.-Y.; Solomon, J. M.; Jiang, J.-K.; Luo, C.-W.; Chang, W.-H.; Her, T.-H. Dielectric breakdown and sub-wavelength patterning of monolayer hexagonal boron nitride using femtosecond pulses. *2D Materials* **2023**, *10* (4). DOI: 10.1088/2053-1583/acfa0f

(33) Wang, X. J.; Fang, H. H.; Li, Z. Z.; Wang, D.; Sun, H. B. Laser manufacturing of spatial resolution approaching quantum limit. *Light Sci Appl* **2024**, *13* (1), 6. DOI: 10.1038/s41377-023-01354-5

(34) Sedghamiz, E.; Liu, M.; Wenzel, W. Challenges and limits of mechanical stability in 3D direct laser writing. *Nat Commun* **2022**, *13* (1), 2115. DOI: 10.1038/s41467-022-29749-9

(35) Luo, H.; Yu, H.; Wen, Y.; Zheng, J.; Wang, X.; Liu, L. Direct Writing of Silicon Oxide Nanopatterns Using Photonic Nanojets. *Photonics* **2021**, *8* (5). DOI: 10.3390/photonics8050152

(36) Andres-Penares, D.; Habil, M. K.; Molina-Sánchez, A.; Zapata-Rodríguez, C. J.; Martínez-Pastor, J. P.; Sánchez-Royo, J. F. Out-of-plane trion emission in monolayer WSe2 revealed by whispering gallery modes of dielectric microresonators. *Communications Materials* **2021**, *2* (1). DOI: 10.1038/s43246-021-00157-8

(37) Yang, L.; Li, L.; Wang, Q.; Xing, C.; Ma, L.; Zeng, Y.; Zhao, Y.; Yan, Y. Over 1000‐Fold Enhancement of the Unidirectional Photoluminescence from a Microsphere‐Cavity‐Array‐Capped QD/PDMS Composite Film for Flexible Lighting and Displays. *Advanced Optical Materials* **2019**, *7* (24). DOI: 10.1002/adom.201901228

(38) Kfir, O.; Lourenco-Martins, H.; Storeck, G.; Sivis, M.; Harvey, T. R.; Kippenberg, T. J.; Feist, A.; Ropers, C. Controlling free electrons with optical whispering-gallery modes. *Nature* **2020**, *582* (7810), 46-49. DOI: 10.1038/s41586-020-2320-y

(39) Shin, D. H.; Yang, X.; Caneva, S. Single-Molecule Protein Fingerprinting with Photonic Hexagonal Boron Nitride Nanopores. *Acc Mater Res* **2023**, *4* (4), 307-310. DOI: 10.1021/accountsmr.3c00016

(40) Yang, H.; Cornaglia, M.; Gijs, M. A. Photonic nanojet array for fast detection of single nanoparticles in a flow. *Nano Lett* **2015**, *15* (3), 1730-1735. DOI: 10.1021/nl5044067

(41) Sanchez, D. A.; Dai, Z.; Lu, N. 2D Material Bubbles: Fabrication, Characterization, and Applications. *Trends in Chemistry* **2021**, *3* (3), 204-217. DOI: 10.1016/j.trechm.2020.12.011

(42) Lee, H. Y.; Sarkar, S.; Reidy, K.; Kumar, A.; Klein, J.; Watanabe, K.; Taniguchi, T.; LeBeau, J. M.; Ross, F. M.; Gradecak, S. Strong and Localized Luminescence from Interface Bubbles Between Stacked hBN Multilayers. *Nat Commun* **2022**, *13* (1), 5000. DOI: 10.1038/s41467-022-32708-z

(43) Khestanova, E.; Guinea, F.; Fumagalli, L.; Geim, A. K.; Grigorieva, I. V. Universal shape and pressure inside bubbles appearing in van der Waals heterostructures. *Nat Commun* **2016**, *7*, 12587. DOI: 10.1038/ncomms12587

(44) Juodkazis, S.; Nishimura, K.; Tanaka, S.; Misawa, H.; Gamaly, E. G.; Luther-Davies, B.; Hallo, L.; Nicolai, P.; Tikhonchuk, V. T. Laser-induced microexplosion confined in the bulk of a sapphire crystal: evidence of multimegabar pressures. *Phys Rev Lett* **2006**, *96* (16), 166101. DOI: 10.1103/PhysRevLett.96.166101

(45) Castelletto, S.; Maksimovic, J.; Katkus, T.; Ohshima, T.; Johnson, B. C.; Juodkazis, S. Color Centers Enabled by Direct Femto-Second Laser Writing in Wide Bandgap Semiconductors. *Nanomaterials (Basel)* **2020**, *11* (1). DOI: 10.3390/nano11010072

(46) Yang, X.; Hong, M. Enhancement of axial resolution and image contrast of a confocal microscope by a microsphere working in noncontact mode. *Appl Opt* **2021**, *60* (17), 5271-5277. DOI: 10.1364/AO.425028

(47) Yan, Y.; Li, L.; Feng, C.; Guo, W.; Lee, S.; Hong, M. Microsphere-Coupled Scanning Laser Confocal Nanoscope for Sub-Diffraction-Limited Imaging at 25 nm Lateral Resolution in the Visible Spectrum. *ACS Nano* **2014**, *8* (2), 1809-1816. DOI: 10.1021/nn406201q

(48) Chen, L.-W.; Zhou, Y.; Wu, M.-X.; Hong, M.-H. Remote-mode microsphere nano-imaging: new boundaries for optical microscopes. *Opto-Electronic Advances* **2018**, *1* (1), 17000101-17000107. DOI: 10.29026/oea.2018.170001

(49) Cardona, M. Optical Properties and Band Structure of $SrTiO_3$ and $BaTiO_3$. *Physical Review* **1965**, *140* (2A), A651-A655. DOI: 10.1103/PhysRev.140.A651

**Graphical TOC Entry**

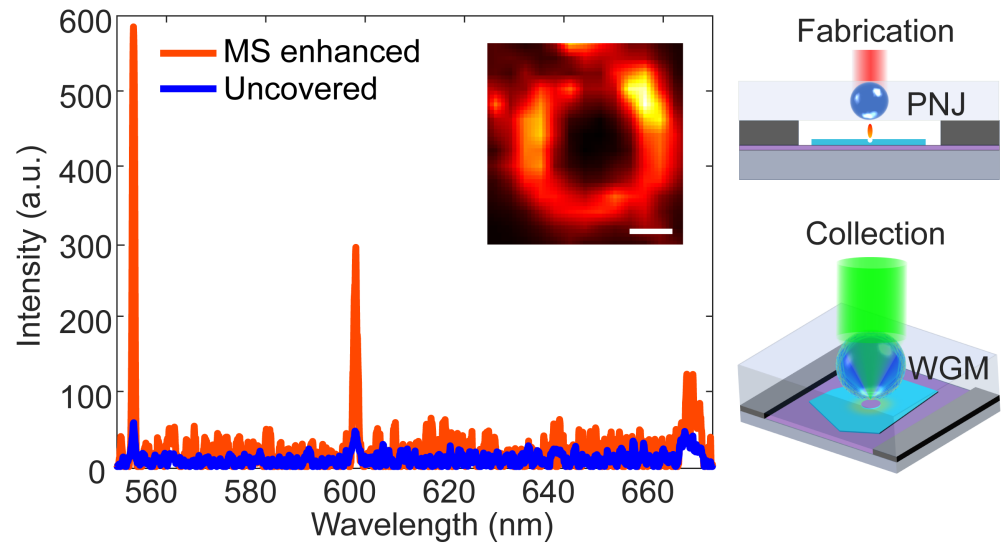

Supporting Information

# Microsphere-assisted generation of localized optical emitters in 2D hexagonal boron nitride

*Xiliang Yang,[†] Dong Hoon Shin,[±] Kenji Watanabe,[§] Takashi Taniguchi,[§] Peter G. Steeneken,[†] Sabina Caneva,[†]\**

The supplementary information includes:

S1. Fabrication of the microsphere-PDMS membrane film.

S2. Height profile along the edge of the tape spacer.

S3. Optical images of defects created by direct fs laser fabrication.

S4. Percentage distribution of ZPL wavelengths with increasing fabrication power.

S5. Simulated electric field distribution for the emitter collection.

S6. Simulated electric field distribution on the microsphere.

S7. Optical images at different image planes of the microsphere.

## S1: Fabrication of microsphere-PDMS membrane film

The experimental procedure to generate the MPM is outlined in Figure S1 below.

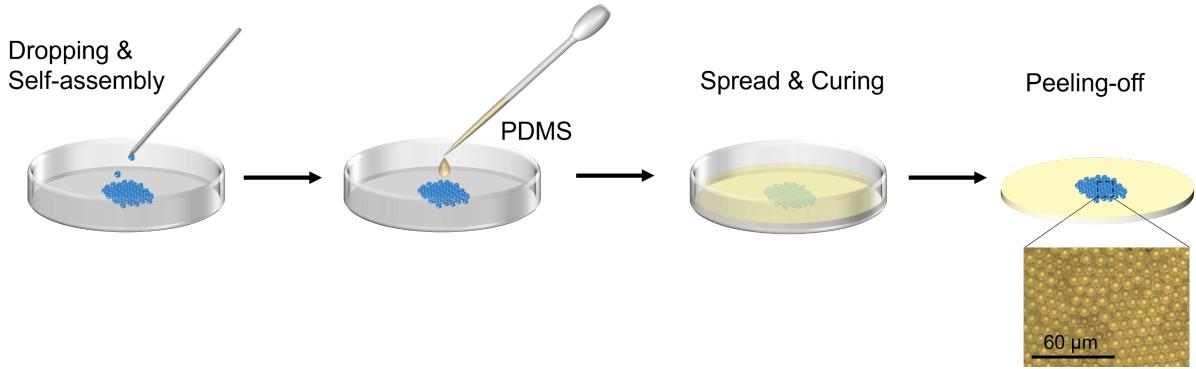

**Figure S1.** Fabrication of microsphere-PDMS membrane film by microsphere dropping and self-assembly, PDMS spreading and curing with microsphere beads, and pealing-off the microsphere film by the tweezer. Insert figure: optical image of microsphere-PDMS membrane film.

## S2. Height profile along the edge of the tape spacer

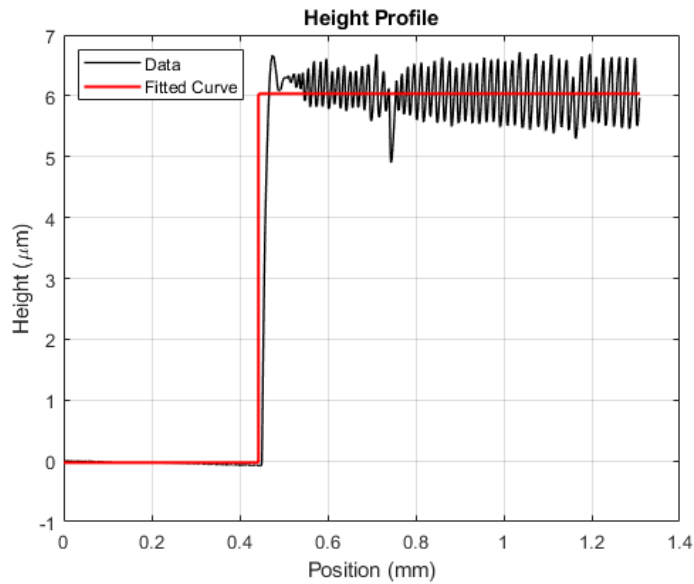

**Figure S2.** Profilometer measurement of the spacer edge. The sticky upper surface of the double-sided tape causes some stickiness in the scanning tip, leading to a less smooth scan result. Despite these variations, the tape's thickness, approximately 6 microns, is still clearly discernible.

## S3. Optical images of defects created by direct fs laser fabrication

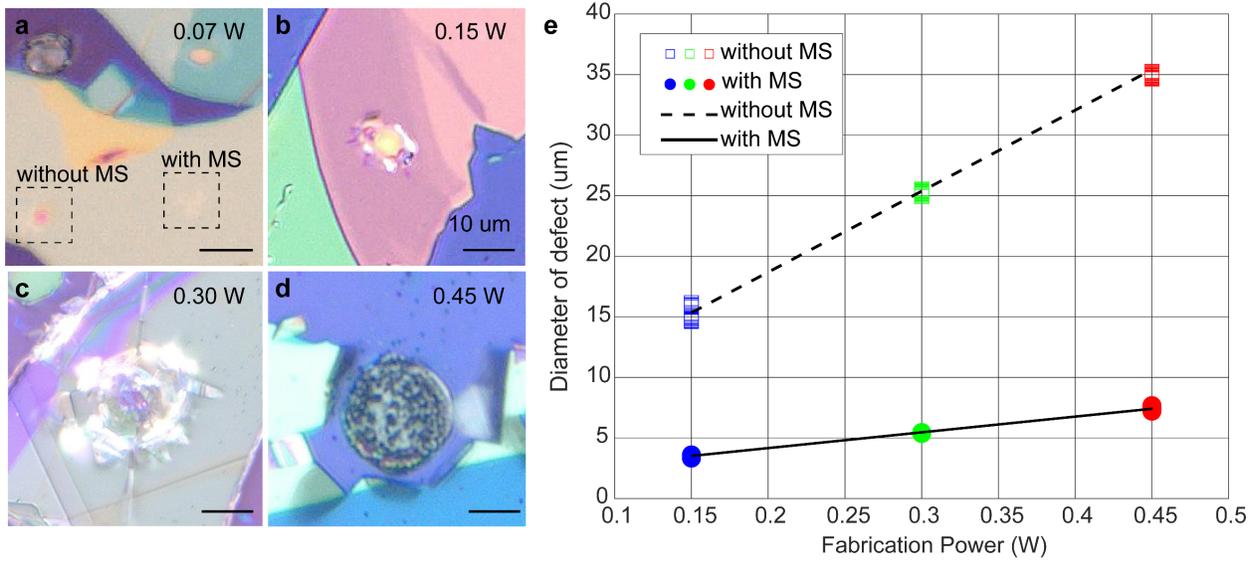

**Figure S3.** Optical images of defects created by direct fs laser fabrication with varying power levels: (a) 0.07 W, (b) 0.15 W, (c) 0.3 W, and (d) 0.45 W. Scale bar: 10 μm. (e) Diameter of defects vs fs laser power for both with and without the microscope stage (MS). Data from 15 or more defects at each power level are included. The dashed line represents a linear fit to the diameter of defects without MS, with a 66.87 μm/W slope. The dashed-dotted line represents a linear fit to the diameter of defects with MS, with a slope of 13 μm/W.

## S4. Percentage distribution of ZPL wavelengths with increasing fabrication power

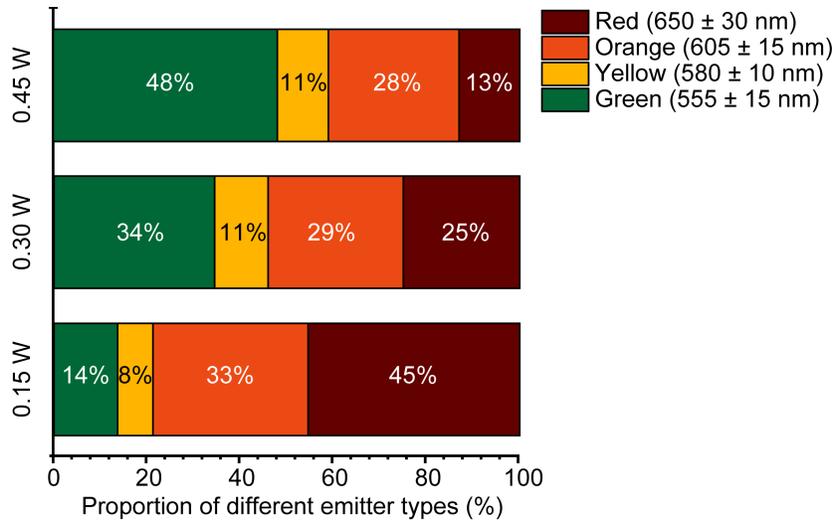

**Figure S4** Percentage distribution of ZPL wavelengths with increasing fabrication power

## S5. Simulated electric field distribution for the emitter collection

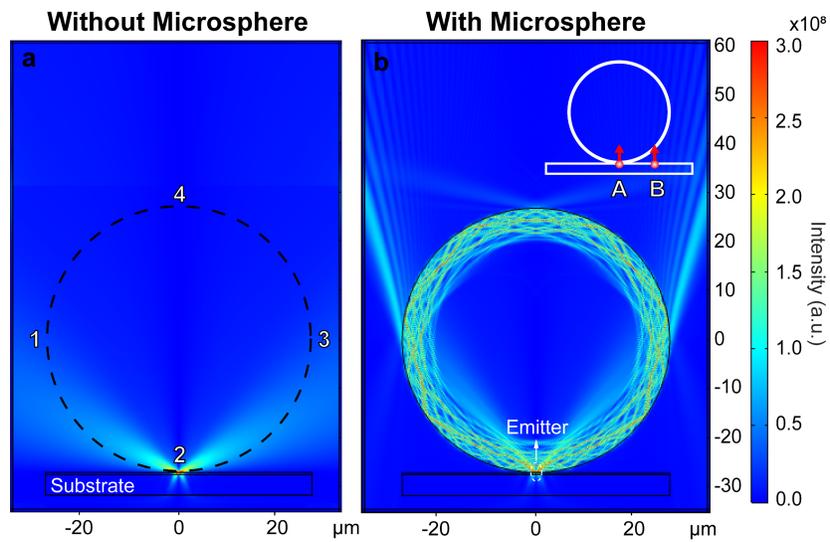

**Figure S5.** Simulated electric field distribution for the emitter collection is shown in (a) without and (b) with the MSM system. The insets display the emitter locations: at the center-bottom of the microsphere and off-center at the bottom of the microsphere, respectively.

## S6. Simulated electric field distribution on the microsphere

**Table S6.** Simulated electric field distribution at points 1,2,3 and 4 on the microsphere and without the microsphere as shown in Figure S5.

| Electric Field (V/m) | Point 1 | Point 2 | Point 3 | Point 4 |
|---|---|---|---|---|
| With MS | $1.10 \times 10^8$ | $1.87 \times 10^{10}$ | $1.10 \times 10^8$ | $1.56 \times 10^8$ |
| Without MS | $2.38 \times 10^7$ | $3.81 \times 10^{10}$ | $2.38 \times 10^7$ | $3.4 \times 10^4$ |

## S7. Optical images at different image planes of the microsphere

Imaging focus on the hBN surface

Imaging through the microsphere

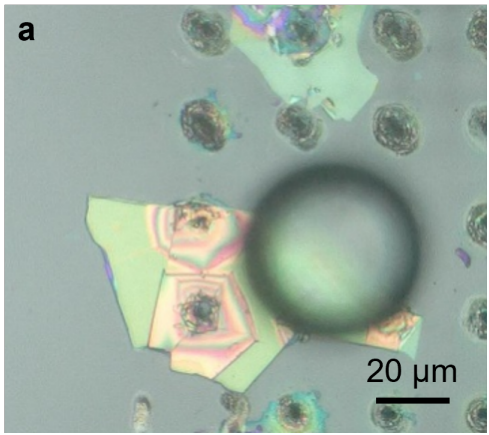
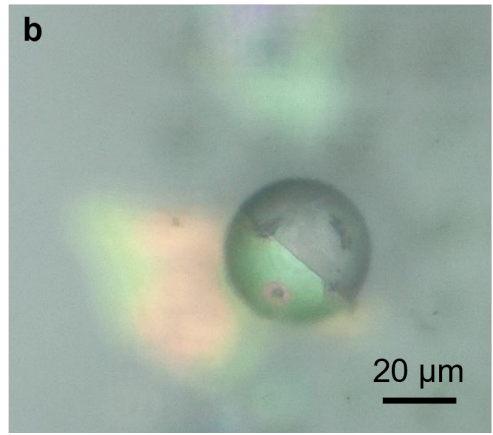

**Figure S7.** Optical images at different image planes of the microsphere. (a) Imaging focused on the hBN surface after fs laser fabrication. (b) Imaging of fabrication defects through the microsphere with virtual image mode.